%% file: ekhara-preprint.tex
\documentclass[preprint,12pt]{elsarticle}

\input{preamble.tex}

\def\ourtitle{
EKHARA Monte Carlo generator for 
      $e^+e^- \to e^+e^- \pi^0$ and
      $e^+e^- \to e^+e^- \pi^+\pi^-$ 
      processes
      }

\begin{document}
\sloppy

\input{frontmatter.tex}

%%%%%%%%%%%%%%%%%%%%%%%%%%%%%%%%%%%%%%%%%%%%%%%%%%%%%%
%%%%%%%%%%%%%%%%%%%%%%%%%%%%%%%%%%%%%%%%%%%%%%%%%%%%%%
%% PROGRAM SUMMARY
%%
\hspace{1pc}{\bf PROGRAM SUMMARY}

\input{prog-summary.tex}

\newpage
%\vspace{30pt}

%%%%%%%%%%%%%%%%%%%%%%%%%%%%%%%%%%%%%%%%%%%%%%%%%%%%%%
%%%%%%%%%%%%%%%%%%%%%%%%%%%%%%%%%%%%%%%%%%%%%%%%%%%%%%
%% LONG WRITE-UP
%%
% In program descriptions the main text of the paper is listed under
% the heading LONG WRITE-UP.
\hspace{1pc}
{\bf LONG WRITE-UP}

%%%%%%%%%%%%%%%%%%%%%%%%%%%%%%%%%%%%%%%%%%%%%%%%%%%%%%
\section{Introduction}
\label{sec:Intro}

\subsection{The physics case}

The first version of EKHARA MC generator~\cite{Czyz:2005ab,Czyz:2006dm}
(EKHARA~ver.~1) was developed to simulate the reaction
 $e^+ e^- \to \pi^+ \pi^- e^+ e^-$. This process is
  a background~\cite{Czyz:2003gb} to the
$e^+ e^- \to \pi^+ \pi^- \gamma (\gamma)$
cross section measurement at meson factories when
 only charged pions are observed~\cite{Aloisio:2004bu,:2008en}. 
 Its proper simulation
is relevant for the precise extraction of the 
 cross section $\sigma(e^+ e^- \to \pi^+ \pi^-)$
and the charged pion form factor
using the {\it radiative return method}.
The description of  
a physics content (matrix elements, modelling of the
 pion-photon interactions, etc.) in EKHARA~ver.~1 was given
  in~\cite{Czyz:2006dm},
however, the computational issues remained
unpublished.
In view of the accepted prolongation
of the experiment, namely,
the KLOE-2 project~\cite{AmelinoCamelia:2010me}, and the planned 
radiative return program of the BES-III experiment \cite{Actis:2010gg,Asner:2008nq},
this gap has to be filled and 
the description of the $e^+ e^- \to \pi^+ \pi^- e^+ e^-$
mode of EKHARA generator is one of the aims 
of this paper.
EKHARA~ver.~1 is also
a convenient base for an inclusion of other channels
of the inelastic $e^+ e^-$ scattering.

The KLOE-2 project assumes an installation of special
tagging devices~\cite{Babusci:2009sg}, which
  will allow~\cite{AmelinoCamelia:2010me} to study, among other,
the $e^+ e^- \to \pi^0 e^+ e^-$ process, 
aiming at the following
measurements:
\begin{itemize}
 \item the two-photon decay width of $\pi^0$,
 \item the transition form factor
       $F_{\pi^0\gamma^*\gamma^*}(m^2_{\pi^0},q_1^2,q_2^2)$
        at space-like photon momentum transfers $q_1^2$, $q_2^2$.
\end{itemize}
These measurements are of a significant 
importance~\cite{AmelinoCamelia:2010me} and the knowledge of
$F_{\pi^0\gamma^*\gamma^*}$
 for high photon virtualities is supposed to help in the reduction of an
 error in
  the calculation of the light-by-light contributions to the muon
  anomalous magnetic moment.
A reliable Monte Carlo generator 
for this channel, based
on the modern knowledge of the $\pi^0-\gamma^*-\gamma^*$  transition form
factor  is an indispensable tool for such studies.
A description of this new mode, which is distributed 
within EKHARA~ver.~2, is the second aim
of this paper.

%%%%%%%%%%%%%%%%%%%%%%%%%%%%%%%%%%%%%%%%%%%%%%%%%%%%%%
\subsection{Basic functionalities of EKHARA~ver.~2}
\begin{itemize}
 \item $e^+e^- \to e^+e^- \pi^+\pi^-$
   \\  -- generates weighted events;
   \\  -- fills  the histograms; 
   \\  -- gives the integrated cross section within cuts.

 \item $e^+e^- \to e^+e^- \pi^0$
   \\  -- generates and stores unweighted events;
   \\  -- fills the histograms; 
   \\  -- gives the integrated cross section within cuts.
\end{itemize}

The program is supplemented with Gnuplot
scripts for visualization of the histograms,
produced by EKHARA.

%%%%%%%%%%%%%%%%%%%%%%%%%%%%%%%%%%%%%%%%%%%%%%%%%%%%%%
\subsection{Related Monte Carlo programs}

There exist several  MC generators 
for $\gamma \gamma$ physics:
  \begin{enumerate}
    \item 
    the code written by A.~Courau~\cite{Courau:1984ia};
    used in~\cite{Alexander:1993rz,Bellucci:1994id} 
    for the $e^+ e^- \to e^+ e^- \pi^0\pi^0$, 
    and in~\cite{Ong:1999gp} 
    for the $e^+ e^- \to e^+ e^- \pi^0$ process; 
    \label{mc:courau}
     \item 
     the code by F.~Nguyen {\it et al.}~\cite{Nguyen:2006sr} 
     for $e^+ e^-\to e^+ e^- \pi^0\pi^0$; 
     \label{mc:npp}
     \item 
     TREPS program written by S.~Uehara~\cite{Uehara:1996di} 
     and used by Belle 
     collaboration~\cite{Mori:2007bu,Mori:2006jj};
     \label{mc:treps}
     \item 
     TWOGAM developed by D.~M.~Coffman,
     used by CLEO~\cite{Gronberg:1997fj} 
     for the $e^+ e^- \to e^+ e^- P$, with 
     $P =\pi^0,\eta,\eta^\prime$;
     \label{mc:twogam}
     \item 
     GGResRC used by the BaBar collaboration~\cite{:2009mc}; 
%     and based on~\cite{Brodsky:1971ud}
%     for single-pseudoscalar production, 
%     and on~\cite{Budnev:1974de} for the two-pion final state; 
     \label{mc:ggresrc}
     \item 
     GALUGA by G.~A.~Schuler~\cite{Schuler:1997ex}
     for LEP2 physics;
     \label{mc:galuga}
     \item 
     GaGaRes written by 
     F.~A.~Berends and R.~van~Gulik~\cite{Berends:2001ta}, 
     for the study of resonance production
     in $\gamma \gamma$ interaction
     at LEP2 energies.
     \label{mc:gagares}
\end{enumerate}
GALUGA and GaGaRes generators describe high-energy physics and 
adapting them to much lower energies is not straightforward.
Other programs are not public.
Moreover, the Equivalent Photon Approximation (EPA)
is employed to a large extent in the majority 
of the generators.
EPA is a useful simplification in the description 
of the two photon processes, when the accuracy requirements are not
high.
It leads however
to some discrepancies with respect to the exact formulation,  
as shown already in~\cite{Brodsky:1971ud}, especially for
single--pseudo\-scalar final states.

Summarizing, the basic requirements for a MC generator for 
$e^+e^- \to e^+e^- P$ 
studies not restricted to the region where the photons 
are quasi--real, are:
\begin{enumerate}
\item use exact formulae and exact kinematics (do not use EPA),
\item include both $s$- and $t$-channel amplitudes and 
       their interference,
\item allow user--defined form factors,
\item implement specific kinematic cuts,
\item account for the peaking behaviour of the cross section,
      in order to have a good Monte Carlo efficiency. 
\end{enumerate}

The EKHARA~ver.~2, presented in this paper, fulfils all the listed
 above criteria.

%%%%%%%%%%%%%%%%%%%%%%%%%%%%%%%%%%%%%%%%%%%%%%%%%%%%%%
\section{The generation of four-momenta for the one-pion mode}
\label{subsec:gener-1pi}

The differential cross-section for the reaction $e^+ e^- \to e^+e^-\pi^0$,
averaged over helicities of the initial $e^+e^-$ states, is given by
\bgea
\label{eq:master-formula-cs-3}
\dd\sigma_{avg}(e^+e^- \to e^+e^-\pi^0) &=& 
\frac{1}{4} \frac{1}{2s} \dLips_3\; \sum{\left| \mathcal{M}_{\pi^0} \right|^2} 
.
\enea
Here $s$ is the initial electron-positron invariant mass squared,
$1/4$ is the averaging factor, $2s$ is the flux factor.
By $\dLips_n$ we denote a differential
$n-$body Lorentz-invariant phase space and $\mathcal{M}_{\pi^0}$
stands for the matrix element describing the reaction $e^+ e^- \to e^+e^-\pi^0$.
We generate the kinematics in the center-of-mass frame of the initial $e^+e^-$,
with the $z$-axis along the initial positron
momentum. 
The event is determined by five kinematic invariants
 and one azimuthal angle~$\phi$:
\bgea
 \dd \sigma(e^+(p_1)\ e^-(p_2) \ \to \ e^+(q_1)\ e^-(q_2)\ \pi^0(Q)) &=& 
 \dd \sigma(s; \ t_1, t_2, s_1, s_2, \phi) ,\nn\\
 s   &=& (p_1 + p_2)^2 ,\nn\\
 t_1 &=& (p_1 - q_1)^2 ,\nn\\
 t_2 &=& (p_2 - q_2)^2 ,\nn\\
 s_1 &=& (p_1 + q_2)^2 ,\nn\\
 s_2 &=& (p_2 + q_1)^2
.
\enea

%%%%%%%%%%%%%%%%%%%%%%%%%%%%%%%%%%%%%%%%%%%%%%%%%%%%%%%%%%%%%%%%%%%%%%%%%
\subsection{The matrix element}

\begin{figure} \begin{center}
 \resizebox{0.29\textwidth}{!}{%
      \includegraphics{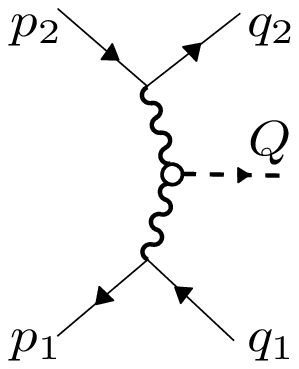} } 
 \hfill
 \resizebox{0.43\textwidth}{!}{%
      \includegraphics{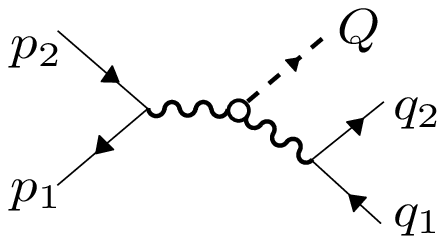} } 
 \end{center}
 \caption{ The $t$-channel {\it (left)} 
           and the $s$-channel {\it (right)} diagrams
           for $e^+ e^- \to e^+ e^- P$}
 \label{fig:diagrams}
 \end{figure}

The matrix element for the reaction $e^+e^- \to e^+e^- P$ 
at the tree level contains the $t$-channel
and the $s$-channel parts
depicted in Fig.~\ref{fig:diagrams} ($\mathcal{M}_{\pi^0}=\mathcal{M}_t+\mathcal{M}_s$).

%%%%%%%%%%%%%%%%%%%%%%%%%%%%%%%%%%%%%%%%%%%%%%%%%%%%%%%%%%%%%%%%%%%%%%%%%
%%%%%%%%%%%%%%%%%%%%%%%%%%%%%%%%%%%%%%%%%%%%%%%%%%%%%%%%%%%%%%%%%%%%%%%%%

The $t$-channel matrix element has the following form
\bgea
\label{eq:msq-t}
 \mathcal{M}_t
 &=& 
 - \; \frac{4\, i \alpha^2}{f_\pi}\;F(t_1, t_2) \;\epsilon_{\mu\nu\alpha\beta}\; 
 \frac{1}{t_1 \; t_2} \; (q_1-p_1)^\alpha \; (q_2-p_2)^\beta 
 \nn\\&&
 \times \; \left( \bar{v}(p_1)\; \gamma^\mu \; v(q_1) \right)
 \; \left(  \bar{u}(q_2) \; \gamma^\nu \; u(p_2) \right)
.
\enea 
The completely antisymmetric tensor $\epsilon^{\mu\nu\alpha\beta}$
is defined by $\epsilon_{0123}= -\epsilon^{0123}=1$.
The $t$-channel contribution to the cross section,
is highly peaked at small  $t_1$ and/or $t_2$ values. 
The pion two-photon transition form factor $F(t_1, t_2)$ 
is an important ingredient, see, e.g.,~\cite{Knecht:2001qf}
and it provides an additional dumping of the amplitude at large values
of $t_1$, $t_2$. 
The normalization is $F(0, 0)=1$, $\alpha\approx 1/137$ is the fine structure
constant and $f_\pi\approx 92.4$~MeV is the pion decay constant. 

\begin{figure} \begin{center}
 \resizebox{0.48\textwidth}{!}{%
      \includegraphics{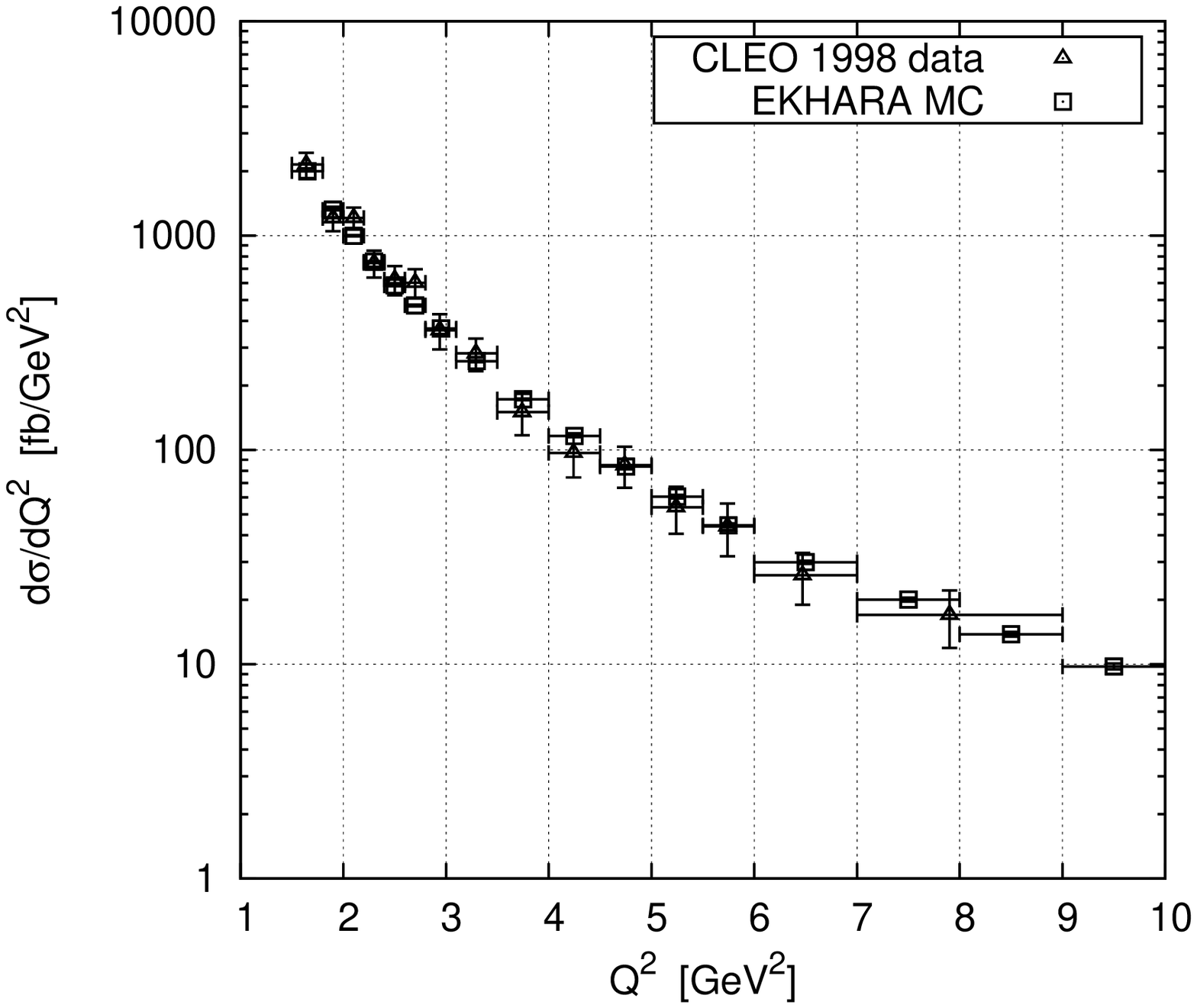} } 
 \hfill
 \resizebox{0.48\textwidth}{!}{%
      \includegraphics{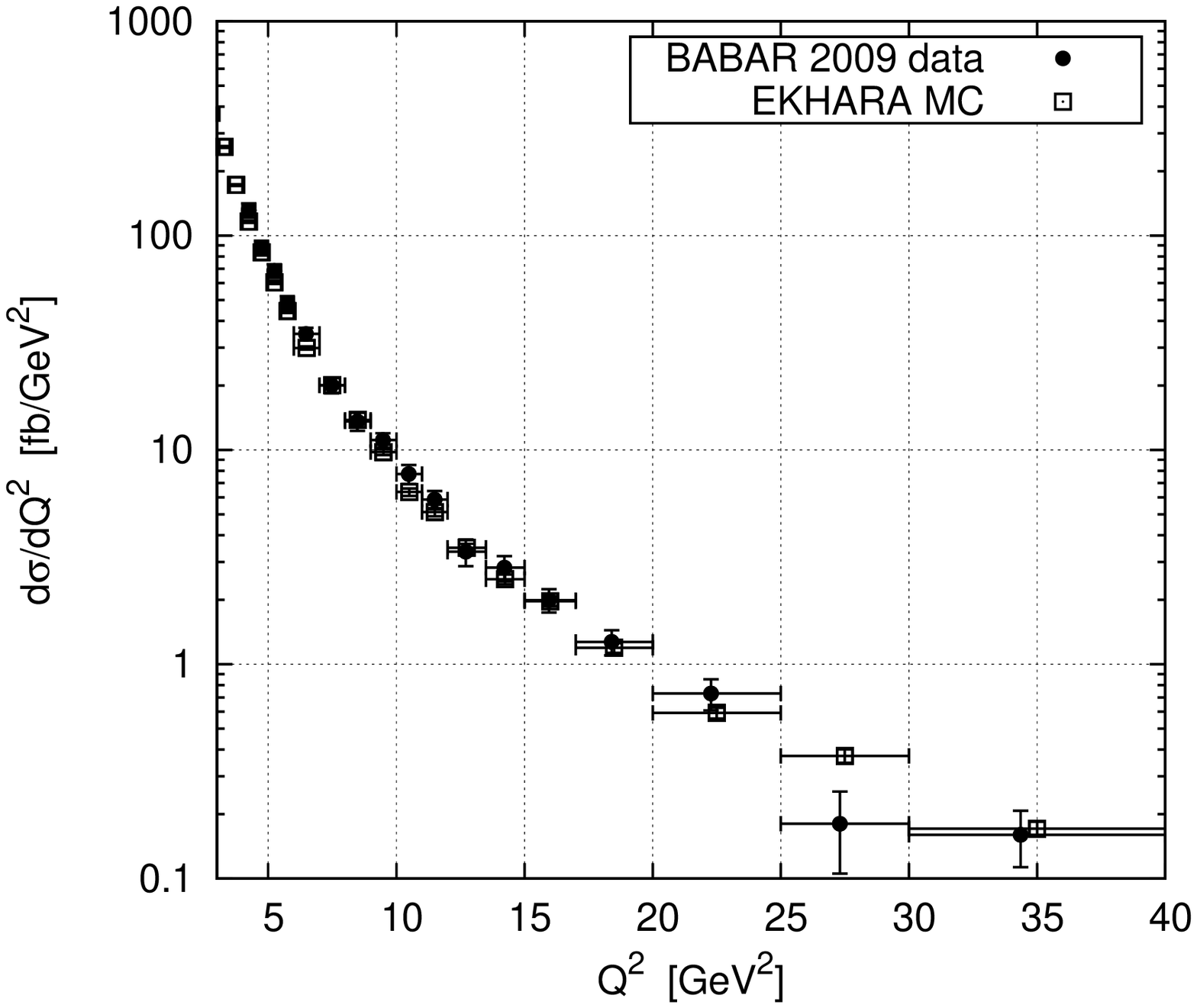} } 
 \end{center}
 \caption{Comparison of EKHARA Monte Carlo 
          simulation with experimental data
          from CLEO~\cite{Gronberg:1997fj}
          and BaBar~\cite{:2009mc} }
 \label{fig:MC-experiment-compar}
 \end{figure}

In Fig.~\ref{fig:MC-experiment-compar}
we demonstrate an agreement of the Monte Carlo simulation with the 
``single-tag'' experimental data
from CLEO~\cite{Gronberg:1997fj} and BaBar~\cite{:2009mc}.
For simulation we use the matrix element~(\ref{eq:msq-t}) with
the form factor of the lowest meson dominance model with two vector resonances 
(LMD+V), fitted~\cite{Nyffeler:2009uw} to the BABAR data~\cite{:2009mc}.
A number of other expressions 
for the form factor are also available for the user of EKHARA. 
For example, one can use the formulae given by the simple vector meson dominance 
ansatz (rho meson pole), or the lowest meson dominance approach results 
with one vector resonance multiplet (LMD) or two multiplets (LMD+V)
with the generic form obtained from the operator product expansion 
considerations~\cite{Knecht:2001qf}. One may also use 
the form factor derived in a quark model, which 
resembles a reasonable agreement with the BaBar~\cite{:2009mc} data
at high momentum transfer~\cite{Dorokhov:2009jd}.

%%%%%%%%%%%%%%%%%%%%%%%%%%%%%%%%%%%%%%%%%%+%%%%%%%%%%%%%%%%%%%%%%%%%%%%%%
%%%%%%%%%%%%%%%%%%%%%%%%%%%%%%%%%%%%%%%%%%+%%%%%%%%%%%%%%%%%%%%%%%%%%%%%%

The matrix element for the $s$-channel reads
\bgea
\label{eq:msq-s}
 \mathcal{M}_s
 &=& 
 \frac{4\, i \alpha^2}{f_\pi}\; F(s, (q_1+q_2)^2) \;\epsilon_{\mu\nu\alpha\beta}\; 
 \frac{1}{s \; (q_1+q_2)^2} \; (p_1+p_2)^\alpha \; (q_1+q_2)^\beta 
 \nn\\&&
 \;\left(   \bar{v}(p_1)\; \gamma^\mu \; u(p_2) \right)
 \;\left(   \bar{u}(q_2) \; \gamma^\nu \; v(q_1)  \right)
.
\enea 
The form factor $F(s, (q_1+q_2)^2)$ in the $s$-channel has to be given by the same 
analytic function as in the $t$-channel case.
The  $s$-channel contribution,
having a peak at small invariant masses of the final $e^+e^-$ pair,
is much smaller than the $t$-channel one 
when at least one of the $t_1$, $t_2$ invariants is small and
it is often missing in the existing MC codes.
In EKHARA all the amplitudes are included
($s$- and $t$-channel as well as their interference). 
This guaranties that no relevant piece is missing
even for the kinematic configurations where both $t_1$ and $t_2$
are large.

The peaking behaviour of the $s-$ and $t-$channel amplitudes
is different.
Therefore EKHARA uses, depending on the set of amplitudes included in the calculation 
($\mathcal{M}_s$, $\mathcal{M}_t$ or both),
 one of the three generation procedures
described below.

%%%%%%%%%%%%%%%%%%%%%%%%%%%%%%%%%%%%%%%%%%%%%%%%%%%%%%%%%%%%%%%%%%%%%%%%%
\subsection{$t-$channel \label{tchannel}}

To generate the particle four-momenta we have adopted
a very efficient generating procedure, which 
was used in the MC generator GALUGA~\cite{Schuler:1997ex}.
Here we only sketch its layout and for further details 
we refer the reader to \cite{Schuler:1997ex}. 
The Lorentz-invariant phase space is mapped to a unit
hypercube in the space of the uniformly distributed 
random numbers $r_i\in[0,1]$, $i=1,\ldots,5$:
\bgea
{\dLips}_3 &=&  
\VV^{t} \, \dd t_1 \dd t_2 \dd s_1 \dd s_2 \dd \phi
= \VV^{t} \, J_t  \, \prod_{i=1}^5 \dd r_i
\ ,
\enea 
where $J_t$ is the mapping Jacobian and
the volume factor reads
\bgea
      \VV^t &=&    \frac{4 \pi^2}{\pi^5\; 2^9}
      \frac{1}{s \beta}
\ ,
\enea
with
\bgea
       \beta  &=& 1 - 4 m_e^2/s
\ .
\enea
The electron mass is denoted by $m_e$.

The user is supposed to provide the following cuts (minima and maxima):
\begin{itemize}
  \item
  for $t_2$: {\tt CUT\_t2min} and {\tt CUT\_t2max},
  \item
  for $t_1$: {\tt CUT\_t1min} and {\tt CUT\_t1max},
  \item
  for positron polar angle: {\tt CUT\_th1min} and {\tt CUT\_th1max},
  \item
  for electron polar angle:  {\tt CUT\_th2min} and {\tt CUT\_th2max},
  \item
  for positron energy:  {\tt E1\_min} and {\tt E1\_max},
  \item
  for electron energy:  {\tt E2\_min} and {\tt E2\_max}.
\end{itemize}
The kinematic invariants 
%$t_1$, $t_2$, $s_1$, $s_2$
are generated in the following sequence:
\begin{enumerate}
 \item calculate cuts on $t_2$
       \begin{itemize}
        \item {\tt t2max} according to {\tt CUT\_th2min} and {\tt E2\_min}, {\tt E2\_max}
        \item {\tt t2min} according to {\tt CUT\_th2max} and {\tt E2\_min}, {\tt E2\_max}
        \item correct {\tt t2max} and {\tt t2min} according to {\tt CUT\_t2min} and {\tt CUT\_t2max}
       \end{itemize}

 \item generate $t_2$; {\tt t2min }$\le t_2 \le${ \tt t2max} %(random number $1$)
 
 \item calculate cuts on $t_1$
       \begin{itemize}
        \item {\tt t1max}  according to {\tt CUT\_th1min} and {\tt E1\_min}, {\tt E1\_max}
              \\ and according to generated $t_2$
              (the lowest is taken)
        \item {\tt t1min}  according to {\tt CUT\_th1max} and {\tt E1\_min}, {\tt E1\_max}
              \\ and according to generated $t_2$
              (the highest is taken)
        \item correct {\tt t1max} and {\tt t1min} according to {\tt CUT\_t1min} and {\tt CUT\_t1max}
       \end{itemize}

 \item generate $t_1$; {\tt t1min }$\le t_1 \le${ \tt t1max} %(random number $2$)

 \item generate $s_1$ %(random number $3$)

 \item generate $s_2$ %(random number $4$)
\end{enumerate}

The changes of variables are the following:
\bgea
  t_2 &=& \mathrm{t2min} \; \exp\left( r_1 \ln\frac{\mathrm{t2max}}{\mathrm{t2min}} \right)
  ,\\
  t_1 &=& \mathrm{t1min} \; \exp\left( r_2 \ln\frac{\mathrm{t1max}}{\mathrm{t1min}} \right)
  ,\\
  s_1 &=& \frac{X_1}{2} + m_e^2 + t_2 + 2 m_e^2\; \frac{t_2}{X_1}
  \ ,\\
  s_2 &=& \frac{- b + \sqrt{\Delta} \sin( \pi (r_4-1/2) )}{2a}  
\ ,
\enea
where
\bgea
  X_1 &=& (\nu + W) (1 + y_1) \exp(r_3 \delta)
      \ ,\\
 \delta &=& \ln\frac{s(1+\beta)^2}{(\nu + W)(1+ y_1)(1+y_2)}
      \ ,\\
       W &=& \sqrt{\nu^2 - t_1\; t_2}
       \ ,\\
       \nu &=& (m_\pi^2 - t_1 - t_2)/2
       \ ,\\
       y_{1,2} &=& \sqrt{1 - 4 m_e^2 / t_{1,2}}
\ ,
\enea
where $m_\pi$ stands for the $\pi^0$ mass.
Let $\Delta_4$ be the $4\times4$ symmetric Gram determinant 
of any four independent vectors formed out of $p_1$, $p_2$, $q_1$, $Q$, $q_2$,
then its expansion in powers of $s_2$ determines the 
coefficients $a$, $b$ and $c$:
$16 \Delta_4 \equiv a s_2^2 + b s_2 + c$, we also use
$\Delta = b^2 - 4 ac$.
For numerically stable forms
of $\Delta$, $a$, $b$ and $c$ we refer to~\cite{Schuler:1997ex}.
From $t_1$, $t_2$, $s_1$, $s_2$ one can calculate the final state particles
four-momenta in a frame where the $x-z$ plane is given by the initial
positron and final $\pi^0$ momenta. 
Subsequently, the `event' is randomly rotated around the $z-$axis~\cite{Schuler:1997ex}
using the $r_5$ random number.

For this mapping one has
\bgea
      J_t &=&  \delta \; \ln(\mathrm{t1max}/\mathrm{t1min})\;     
               \ln(\mathrm{t2max}/\mathrm{t2min})\; D_t
          \equiv  \tilde{J_t} \; D_t
     \ ,\\
      D_t  &=& t_1 \; t_2
\ .
\enea
The $D_t$ factor absorbs the peaking behaviour of the $t-$channel amplitude.
The value of $\tilde{J_t} \equiv J_t/D_t$ 
for each given event is stored in the variable
{\tt JacobianFactor}.
The differential cross section with the 
$t-$channel mapping reads
\bgea
\label{eq:master-formula-adapted-t}
\dd\sigma_{avg} &=& \frac{1}{4} \frac{1}{2s} \VV^{t} \, \tilde{J_t}  \, \prod_{i=1}^5 \dd r_i\; 
\left(D_t\,\sum{\left|\mathcal{M} \right|^2} \right)
.
\enea

The above procedure is used when user requires to use the $\mathcal{M}_t$
amplitude only.

%%%%%%%%%%%%%%%%%%%%%%%%%%%%%%%%%%%%%%%%%%%%%%%%%%%%%%
\subsection{$s-$channel \label{schannel}}

The Lorentz-invariant phase space $\dLips_3$ 
in eq.~(\ref{eq:master-formula-cs-3}) can be 
factorized in the following way:
\bgea
\label{eq:dLips3-s}
\dLips_3 &=& \frac{\dd k^2}{2\pi} \dLips_2^i \dLips_2^f
\ ,
\enea
where $k^2$ has a natural meaning
of the invariant mass squared of the virtual photon,
which couples to the final $e^+e^-$ in the 
$s-$channel diagram.

Explicit expressions for these $2$-body phase spaces are
\bgea
\dLips_2^i &=& \frac{1}{(2\pi)^2} \frac{1}{4\sqrt{s}}\sqrt{\frac{(s+k^2-m_\pi^2)^2}{4s} - k^2}\; \dd \Omega_i
\ ,
\\
\dLips_2^f &=& \frac{1}{(2\pi)^2} \frac{1}{4\sqrt{k^2}}\sqrt{\frac{k^2}{4}-m_e^2}\; \dd \Omega_f^\ast 
\ ,
\enea
where
$\dd\Omega_f^\ast$ is the differential solid angle in the center of mass frame of final $e^+e^-$,
i.e. the self frame of the $k^2$ and $\dd \Omega_i$ denotes the pion
solid angle in the center of mass frame of the initial $e^+e^-$ pair.
Thus, (\ref{eq:dLips3-s}) takes the form
\bgea
\label{eq:dLips3-s-simple}
\dLips_3 &=& \VV^{s} \; \dd k^2 \dd\Omega_i \dd\Omega_f^\ast
\ ,
\enea
where the volume factor reads
\bgea
\label{eq:dLips3-s-phvol}
\VV^{s}&=& 
\frac{1}{2^9\pi^5}
\frac{1}{\sqrt{sk^2}}
\sqrt{\frac{k^2}{4}-m_e^2} 
\sqrt{\frac{(s+k^2-m_\pi^2)^2}{4s} - k^2}
\ .
\enea

The Lorentz-invariant phase space is mapped to a unit
hypercube in the space of the uniformly distributed 
random numbers $r_i\in[0,1]$, $i=1,\ldots,5$:
\bgea
{\dLips}_3 &=& \VV^{s} \, J_s \, \prod_{i=1}^5 \dd r_i \ ,
\enea 
where $J_s$ is the Jacobian for a mapping from 
physical variables to $r_i$.

The following mapping is adopted:
\bgea
\label{eq:r_i-definition}
\phi_i &=& 2\pi r_1
\ ,
\\
\nn
\cos \theta_i &=&  -1 + 2 r_2;\; \sin \theta_i = 2\sqrt{r_2(1-r_2)}
\ ,
\\
\nn
\phi_f^\ast &=& 2\pi r_3
\ ,
\\
\nn
\cos \theta_f^\ast &=&  -1 + 2 r_4;\; \sin \theta_i^\ast = 2\sqrt{r_4(1-r_4)}
\ ,
\enea 
and the invariant mass squared ($k^2$) of the final $e^+e^-$ pair
is generated using logarithmic mapping
\bgea
k^2 &=& 4m_e^2 
\left(
\frac{(\sqrt{s}-m_\pi)^2}{4m_e^2} 
\right)^{r_5}
.
\label{eq:kk-definition}
\enea

From~(\ref{eq:r_i-definition}) and~(\ref{eq:kk-definition})
one can recover the momenta of all final state particles:
\bgea
|\vec{Q}| &=& \frac{1}{2\sqrt{s}} \sqrt{\lambda(s,k^2,m_\pi^2)}
\ ,
\\
\nn
\vec{Q} &=& |\vec{Q}| \, 
\left(
\begin{array}{c}
\sin\theta_i\, \sin\phi_i
\\
\sin\theta_i\, \cos\phi_i
\\
\cos\theta_i
\end{array}
\right)
,
\\
\nn
k &\equiv& (E_k, \vec{k}) =( \sqrt{k^2 + |\vec{Q}|^2} , \,  -\vec{Q})
\ ,
\enea
and
\bgea
|\vec{q_2}^\ast| &=& \frac{1}{2\sqrt{k^2}} \sqrt{\lambda(k^2,m_e^2,m_e^2)}
\ ,
\\
\nn
\vec{q_2}^\ast &=& |\vec{q_2}^\ast| \, 
\left(
\begin{array}{c}
\sin\theta_f^\ast\, \sin\phi_f^\ast
\\
\sin\theta_f^\ast\, \cos\phi_f^\ast
\\
\cos\theta_f^\ast
\end{array}
\right)
,
\\
\nn
{q_2}^\ast &=& ( \sqrt{m_e^2 + |\vec{q_2}^\ast|^2} , \, \vec{q_2}^\ast)
\ ,
\\
\nn
{q_1}^\ast &=& ( \sqrt{m_e^2 + |\vec{q_2}^\ast|^2} , \, -\vec{q_2}^\ast)
\ .
\enea

The kinematic variables with asterisks
are given in the  center of mass frame of the final $e^+e^-$ pair,
i.e. in the self frame of the second virtual photon.
In order to obtain the values of the final positron and electron
momenta in the lab frame,
the $q_1$ and $q_2$,
a Lorentz transformation of $q_1^\ast$ and $q_2^\ast$ is performed.
%%%%%%%%%%%%%%%%%
%The following formulae are used:
%\bgea
%\label{eq:boost-s}
%(q_i)^0 &=&
%\gamma 
%\left(
%(q_i^\ast)^0 + \vec{q_i}^\ast \vectimes \vec{k} / E_k
%\right)
%,
%\\
%\nn
%\vec{q_i} &=& \vec{q_i}^\ast + \vec{k} 
%\left(
%(q_i^\ast)^0\,\gamma/E_k +  \vec{q_i}^\ast \vectimes \vec{k} \,(\gamma -1)/|\vec{k}|^2
%\right)
%,
%\\
%\nn
%\gamma &=& E_k/\sqrt{k^2}
%,
%\;\;\;\;\;\;\;\;\;i=1,\,2
%.
%\enea
%%%%%%%%%%%%%%%%%

For this mapping, we have
\bgea
J_s &=& 
4(2\pi)^2 \, 
\ln\left( \frac{(\sqrt{s}-m_\pi)^2}{4m_e^2} \right)
\; D_s \equiv \tilde{J}_s \; D_s
\ ,
\\
D_s &=& {k^2}
\ .
\enea
The $D_s$ factor absorbs the peaking behaviour of the $s-$channel amplitude.
The value of  $\tilde{J_s} \equiv J_s/D_s$ 
for each given event is stored in the variable
{\tt JacobianFactor}.
The differential cross section with the 
$s-$channel mapping reads
\bgea
\label{eq:master-formula-adapted-s}
\dd\sigma_{avg} &=& \frac{1}{4} \frac{1}{2s} \VV^{s} \, \tilde{J_s}  \, \prod_{i=1}^5 \dd r_i\; 
\left(D_s\,\sum{\left|\mathcal{M} \right|^2} \right)
.
\enea

The above procedure is used when user requires to use the $\mathcal{M}_s$
amplitude only.

%%%%%%%%%%%%%%%%%%%%%%%%%%%%%%%%%%%%%%%%%%%%%%%%%%%%%%
\subsection{Merging $s-$ and $t-$channels}

The differential cross section can be written as
\bgea
\label{eq:master-formula-adapted-s_and_t}
\dd\sigma_{avg} &=& \frac{1}{4} \frac{1}{2s} 
\prod_{i=1}^5 \dd r_i \, \VV \, J 
           \left( \frac{A/D_s + B/D_t}{A/D_s + B/D_t}  \right)
\, \sum{\left|\mathcal{M} \right|^2}
\\
&=& \frac{1}{4} \frac{1}{2s} 
\prod_{i=1}^5 \dd r_i  \, 
           \left( A\; \VV^s \, \tilde{J}_s + B\; \VV^t \, \tilde{J}_t  \right)
\frac{D_s D_t}{B D_s + A D_t} 
\, \sum{\left|\mathcal{M} \right|^2}
\\
\nn
&=&
\quad\ \ \frac{1}{4} \frac{1}{2s} 
\int_0^A \dd r_6
\prod_{i=1}^5 \dd r_i  \, \VV^s \, \tilde{J}_s 
\frac{D_s D_t}{B D_s + A D_t} 
\, \sum{\left|\mathcal{M} \right|^2}
\\
&& +
\quad \frac{1}{4} \frac{1}{2s} 
\int_A^1 \dd r_6
\prod_{i=1}^5 \dd r_i  \,  \VV^t \, \tilde{J}_t
\frac{D_s D_t}{B D_s + A D_t} 
\, \sum{\left|\mathcal{M} \right|^2}
\ ,
\enea
where $r_i\in[0,1]$, $i=1,\ldots,5$ are 
the uniform random numbers,
and $A\equiv(1-r_m)$, $B\equiv r_m$, with an a priori weight $r_m \in (0,1)$.
We find empirically that $r_m= 0.9$ gives an 
efficient merging in order to 
generate events distributed according to
$|M_s + M_t|^2$. The s-channel (t-channel) is used with a probability
 $1-r_m$ ($r_m$). In each channel procedures described in Sections
 \ref{tchannel} and \ref{schannel}) are used.

The above procedure is followed when user requires to use the 
full amplitude $\mathcal{M} = \mathcal{M}_s + \mathcal{M}_t$.

%%%%%%%%%%%%%%%%%%%%%%%%%%%%%%%%%%%%%%%%%%%%%%%%%%%%%%
\subsection{Generation of the unweighted events}
Let the $UB$ be a pre-evaluated upper bound for 
the Monte Carlo integrand $Contrib$.
For generation of the unweighted events,
we use the following accept--reject method:
\begin{enumerate}
 \item 
     calculate $Contrib$,
 \item 
      generate random number $r_{accept}$,
 \item 
      accept event, if $Contrib \ge r_{accept} \, UB$.
\end{enumerate}
The explicit expression for $Contrib$ depends on the
mapping, which is used by the generator, and reads
\begin{itemize}
  \item $s-$channel
     \bgea
     \nn
      Contrib &=& \frac{1}{4} \VV^s \, J_s 
                  \, \sum{\left|\mathcal{M} \right|^2}
      ,
     \enea
  \item $t-$channel
     \bgea
     \nn
      Contrib &=& \frac{1}{4} \VV^t \, J_t 
                  \, \sum{\left|\mathcal{M} \right|^2}
      ,
     \enea
  \item Merged $s-$ and $t-$channel
     \bgea
     \nn
      Contrib &=& \left\{ \begin{array}{ll}
                 \frac{1}{4} \VV^s \, \tilde{J}_s 
                 \frac{D_s D_t}{B D_s + A D_t} 
                  \, \sum{\left|\mathcal{M} \right|^2}
                 & with \ probability \ (1-r_m) , \\
                 &   \\
                 \frac{1}{4} \VV^t \, \tilde{J}_t 
                 \frac{D_s D_t}{B D_s + A D_t} 
                  \, \sum{\left|\mathcal{M} \right|^2}
                 & with \ probability \ (r_m) .
                  \end{array} \right.
     \enea
\end{itemize}

\begin{figure} \begin{center}
 \resizebox{0.96\textwidth}{!}{%
      \includegraphics{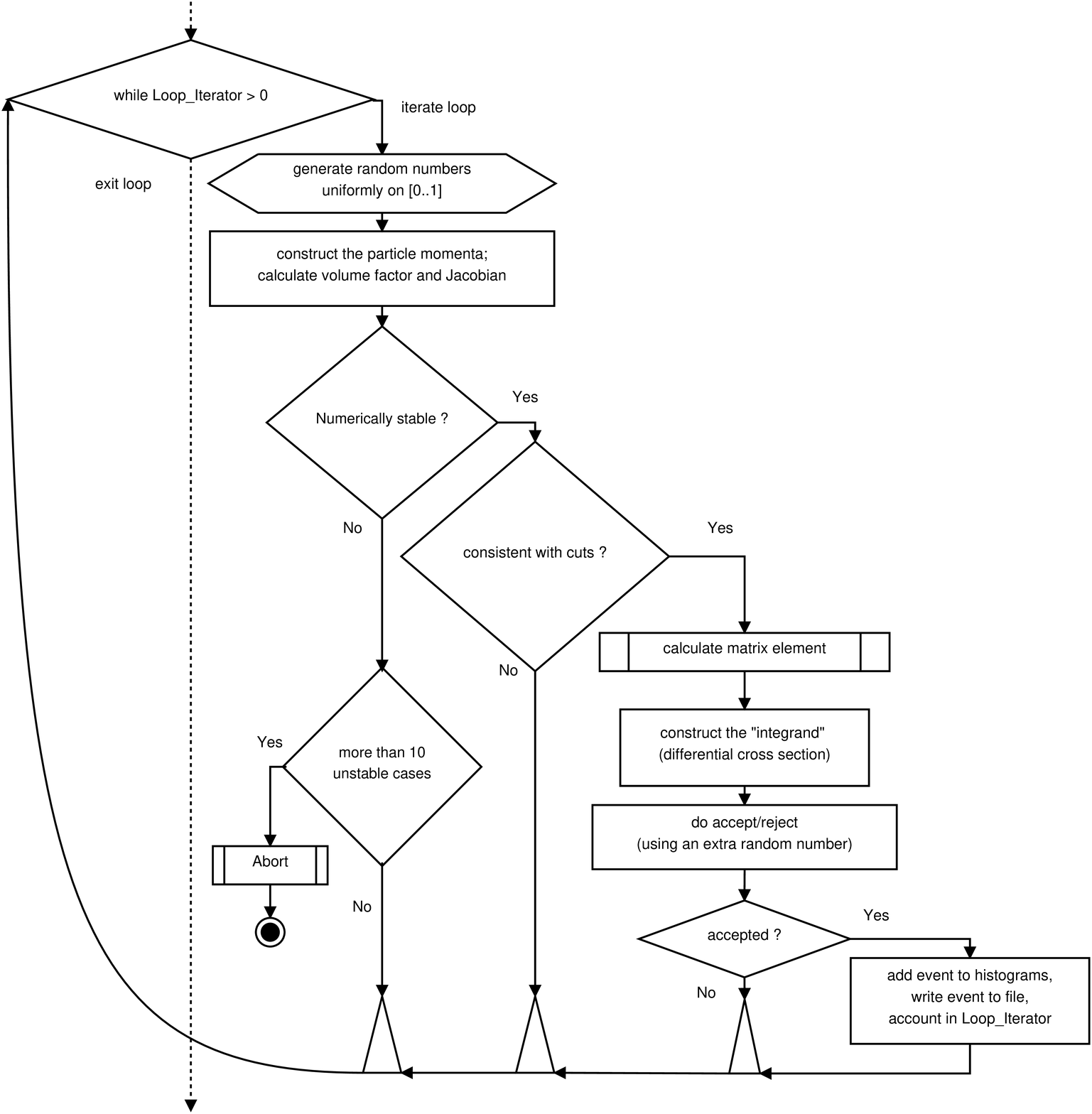} } 
 \end{center}
 \caption{ Flowchart for the $e^+e^- \to e^+e^- \pi^0$ event generation.
           %{\tt SUBROUTINE mc\_loop\_1pi} (mode $e^+e^- \to e^+e^- \pi^0$).
         }
 \label{fig:chart-3}
 \end{figure}

The flowchart for Monte Carlo generator routine is
given in Figure~\ref{fig:chart-3}. It illustrates 
the order of phase space generation, application of cuts
and the use of accept/reject method.
In the block of the numerical stability control
we make sure that the upper bound of the integrand is
not overshooted, the energy and momentum conservation
holds true, the final particles have on-shell momenta,
the Gramm determinants are positively defined
and that the generated sines and cosines of 
the particle angles are within $[-1,1]$.

%%%%%%%%%%%%%%%%%%%%%%%%%%%%%%%%%%%%%%%%%%%%%%%%%%%%%%%%%%%%%%%%%%%%%%%%%
%%%%%%%%%%%%%%%%%%%%%%%%%%%%%%%%%%%%%%%%%%%%%%%%%%%%%%%%%%%%%%%%%%%%%%%%%

%%%%%%%%%%%%%%%%%%%%%%%%%%%%%%%%%%%%%%%%%%%%%%%%%%%%%%
\section{The generation of four-momenta in the two-pions mode}
\label{subsec:gener-2pi}

The differential cross-section for the reaction $e^+(p_1)e^-(p_2) \to \pi^+(\pi_1)\pi^-(\pi_2)e^+(q_1)e^-(q_2)$,
averaged over helicity states of the initial $e^+e^-$, is given by
\bgea
\label{eq:master-formula-cs-4}
\dd\sigma_{avg}(e^+ e^- \to \pi^+ \pi^- e^+ e^-) &=& \frac{1}{4} \frac{1}{2s} \dLips_4\; \sum{\left| \mathcal{M}_{\pi^+ \pi^-} \right|^2} 
.
\enea

The explicit expression for the matrix element $\mathcal{M}_{\pi^+ \pi^-}$
can be found in~\cite{Czyz:2006dm} and
its numerical evaluation is carried out
in the helicity amplitudes framework.

The multi--channel variance reduction method is used
to improve efficiency of the
 generator and the generation is split into four channels, where two of them
 absorb peaks present in $t-$channel diagrams and other two take care
 of the $s-$channel peaks. At difference to one-pion mode only one
 procedure for generation is used independently on the included
 contributions.
 This guaranties an efficient generation when s- and t-channel
 contributions are summed in the matrix element. Moreover, as this mode
 was constructed for generation of a background for the radiative return
 events, the generator is not optimised to run in the region, where
 $\gamma^*\gamma^*$ contributions dominate.

In this Section we use the following definitions:
\begin{equation}
 \ k_1=q_1+q_2\ ,
 \ Q=\pi_1+\pi_2\ , 
\end{equation}
\begin{equation}
s=(p_1+p_2)^2\ , \ t=(q_2- p_2)^2 \ , t_1=(p_1- q_1)^2 .
\end{equation}

%%%%%%%%%%%%%%%%%%%%%%%%%%%%%%%%%%%%%%%%%%%%%%%%%%%%%%
\subsection{$t-$channel\label{pi2t}}
 For the $t-$channel peaks absorption, we use the following
 phase space representation  
\bgea
&&\kern-30pt{\dLips}_4(p_1+p_2;q_1,q_2,\pi_1,\pi_2) = \nn \\
 &&\kern-20pt{\dLips}_2(p_1+p_2;Q',q_2)
     \frac{dQ'^2}{2\pi}{\dLips}_2(Q';Q,q_1)
        \frac{dQ^2}{2\pi}{\dLips}_2(Q;\pi_1,\pi_2)
\enea
in one of the channels and an analogous one, with $q_1\leftrightarrow q_2$,
 in the other channel. As both channels are completely symmetric under
  $q_1\leftrightarrow q_2$, we will describe here only changes of variables,
 which smoothen the distribution, only in one of them.
 For the two invariant masses ($Q^2$ and $Q'^2$) the
  following change of variables was performed 
\bgea
\kern-20ptQ^2 =  \left(\sqrt{s}-2m_e-\sqrt{-2z}\right)^2 , \ \ 
z = -\frac{1}{2}(\sqrt{s}-2m_e-2m_\pi)^2(1-r_{Q^2}) \ ,
\enea
 \bgea
\kern-20ptQ'^2 &=& \frac{1}{^3\sqrt{-3y}}+m_e^2\ ,  \\
\kern-20pty &=&-\frac{1}{3(Q^2+2\sqrt{Q^2}m_e)^3}
 +\left(\frac{1}{3(Q^2+2\sqrt{Q^2}m_e)^3}
 -\frac{1}{3(s-2\sqrt{s}m_e)^3}\right)r_{Q'^2} \ . \nn
\enea
 The angles of $\vec q_2$ vector are defined in the initial $e^+e^-$ center
 of mass (cms) frame with z-axis along $\vec p_1$ 
 and the polar angle is used to absorb the peak coming
 from the propagator of the photon exchanged in the $t$-channel
\bgea
\cos\theta_{q_2} &=& \frac{3m_e^2-s+Q'^2-2t}
{\sqrt{1-\frac{4m_e^2}{s}}\lambda^{1/2}(s,Q'^2,m_e^2)} \ , \
t = -\frac{1}{y}\ , \nn \\
y &=& -\frac{1}{t_{-}} +\frac{s\sqrt{1-\frac{4m_e^2}{s}}
\lambda^{1/2}(s,Q'^2,m_e^2)}{m_e^2(Q'^2-m_e^2)^2}r_{\theta_{q_2}} \ , \
\phi_{q_2} = 2\pi r_{\phi_{q_2}} \ ,
\enea
where
\begin{equation}
t_{-} = \frac{1}{2}\left(3m_e^2-s+Q'^2-\sqrt{1-\frac{4m_e^2}{s}}
\lambda^{1/2}(s,Q'^2,m_e^2)\right) \ .
\end{equation}

The angles of the $\vec Q$ vector are defined 
in $Q'$ rest frame and the appropriate change of variables reads 
\bgea
\cos\theta_Q &=& \frac{2E'_{1}Q_0-Q^2-\frac{1}{2|\vec{p'_1}||\vec{Q}|x}}
{2|\vec{p'_1}||\vec{Q}|}\ , \ \phi_{Q} = 2\pi r_{\phi_{Q}}  \\
 &&\kern-50pt x = \frac{-1}{2|\vec{p'_1}||\vec{Q}|(Q^2-2E'_{1}Q_0-2|\vec{p'_1}||\vec{Q}|)}
+ \frac{2}{(Q^2-2E'_{1}Q_0)^2-4|\vec{p'_1}|^2|\vec{Q}|^2}r_{\theta_Q}
 \ , \nn
\enea
where $ E'_{1}= \frac{Q'^2-t+m_e^2}{2\sqrt{Q'^2}}$,
$Q_0 = \frac{Q'^2+Q^2-m_e^2}{2\sqrt{Q'^2}}$,
$|\vec{p'_1}| = \frac{\lambda^{1/2}(Q'^2,t,m_e^2)}{2\sqrt{Q'^2}}$,
$|\vec{Q}| = \frac{\lambda^{1/2}(Q'^2,Q^2,m_e^2)}{2\sqrt{Q'^2}}$ and
 $\lambda(a,b,c)=a^2+b^2+c^2-2(ab+ac+bc)$. As we chose here
 the z-axis along the $\vec{p'_1}$  (the $\vec p_1$ in the $Q'$ rest frame)
 the vectors are rotated
 after generation to restore the general choice of the z-axis
 along $\vec p_1$ in the $e^+e^-$ cms frame.

 Finally the angles of the positively charged pion are generated
 in the $Q$ rest frame with flat distributions

 \bgea
\cos\theta_{\pi_1} = -1+ 2r_{\theta_{\pi_1}} \ , \ 
 \phi_{\pi_1} = 2\pi r_{\phi_{\pi_1}} \ .
\enea

The described change of variables
 transforms the phase space into a unit hypercube 
 ($0<r_i<1\ , \ i= Q^2,\cdots , \phi_{\pi_1} $) and collecting
 all the jacobians it reads

\bgea
&&\kern-20pt{\dLips}_4(p_1+p_2;q_1,q_2,\pi_1,\pi_2) = P(q_1,q_2)
dr_{Q^2}dr_{Q'^2}dr_{\theta_{q_2}}dr_{\phi_{q_2}}dr_{\theta_Q}dr_{\phi_Q}
dr_{\theta_{\pi_1}}dr_{\phi_{\pi_1}}\nn\\
\enea
with
\bgea
 P(q_1,q_2)&=& 
 \frac{1}{6(4\pi)^5Q'^2m_e^2} \ \lambda^{1/2}(Q'^2,Q^2,m_e^2) \
\lambda^{1/2}(s,Q'^2,m_e^2) \ \sqrt{1-\frac{4m_{\pi}^2}{Q^2}}
 \nn \\   
&&
\frac{t^2\ (Q'^2-m_e^2)^2 \ (Q^2-2Q\cdot p_1)^2 }
 {(Q^2-2E'_{1}Q_0)^2-4|\vec{p'_1}|^2|\vec{Q}|^2} \ 
 \frac{\sqrt{Q^2}(\sqrt{s}-2m_e-2m_{\pi})^2}{\sqrt{s}-2m_e-\sqrt{Q^2}}
\nn \\ 
&&\left(\frac{1}{(Q^2+2\sqrt{Q^2}m_e)^3}-\frac{1}{(s-2\sqrt{s}m_e)^3}\right)
\ .
\enea

%%%%%%%%%%%%%%%%%%%%%%%%%%%%%%%%%%%%%%%%%%%%%%%%%%%%%%
\subsection{$s-$channel\label{pi2s}}
For the $s$-channel generation it is convenient to write the phase space
 in the following form 
 
 \bgea
&&\kern-30pt{\dLips}_4(p_1+p_2;q_1,q_2,\pi_1,\pi_2) = \nn \\
 &&\kern-15pt
 {\dLips}_2(p_1+p_2;Q,k_1)\frac{dk_1^2}{2\pi}{\dLips}_2(k_1;q_1,q_2)
        \frac{dQ^2}{2\pi}{\dLips}_2(Q;\pi_1,\pi_2) .
\enea

 The two generation channels used here differ only in the generation
 of the electron--positron pair
 invariant mass $k_1^2$ and the change of variables will be
 described simultaneously. The invariant mass $Q^2$ is generated
 with a flat distribution

\begin{equation}
Q^2 = 4m_{\pi}^2 + ((\sqrt{s}-2m_e)^2-4m_{\pi}^2) r_{Q^2}\ .
\end{equation}

Reflecting two leading $k_1^2$ behaviors of the cross section,
the two distinct changes of variables are done in the generation of $k_1^2$:
\bgea
k_1^2 &=& s\exp(y_I^{1/3})\ ,
 \label{log1}
 \\
y_I &=& \ln^3(4m_e^2/s)
 +\left[\ln^3\left(\left(1-\sqrt{Q^2/s}\right)^2\right)
 -\ln^3(4m_e^2/s)\right] 
r_{k_1^2,I} \nn
\enea
\bgea
k_1^2 &=& s\left(1-\exp(-y_{II} ) \right)\ ,
 \label{log2}
 \\
y_{II} &=& - \ln(1-4m_e^2/s) 
 -\ln\left(\frac{\sqrt{Q^2}\left(2\sqrt{s}-\sqrt{Q^2}\right)}
 {(s-4m_e^2)}\right) r_{k_1^2,II} \ . \nn
\enea

 The $\vec k_1$ polar angle is used to absorb peaks coming
 from the electron propagator, while its azimuthal angle
 is generated with a flat distribution:
 
\bgea
\phi_{k_1} &=& 2\pi r_{\phi_{k_1}} \ , \ 
 \cos\theta_{k_1} = \frac{-k_1^2+2k_{10}p_{10}}{2|\vec{k_1}||\vec{p_1}|}
\tanh\left(\frac{y}{2}\right)\nn \\
 &&\kern-70pt y = \ln \left( \frac
{ k_1^2-2k_{10}p_{10} +{2|\vec{k_1}||\vec{p_1}|}}
{ k_1^2-2k_{10}p_{10} -{2|\vec{k_1}||\vec{p_1}|}} \right)
+ \ln \left( \frac
  { k_1^2-2k_{10}p_{10} -{2|\vec{k_1}||\vec{p_1}|}}
{ k_1^2-2k_{10}p_{10} +{2|\vec{k_1}||\vec{p_1}|}}\right)^2
r_{\theta_{k_1}}  ,
\enea
where $k_{10}= \frac{s+k_1^2-Q^2}{2\sqrt{s}}$,
$p_{10} = \frac{\sqrt{s}}{2}  $,
$|\vec{p_1}| = \sqrt{\frac{s}{4}-m_e^2}$ and
$|\vec{k_1}| = \frac{\lambda^{1/2}(s,k_1^2,Q^2)}{2\sqrt{s}}$, 
are defined in the $p_1+p_2$ rest frame.

 The $\vec q_1$ and the $\vec\pi_1$ angles are generated with flat 
 distributions 

\bgea
\kern-10pt
 \phi_{q_1} = 2\pi r_{\phi_{q_1}}, \cos\theta_{q_1} = -1+ 2r_{\theta_{q_1}},
 \phi_{\pi_1} = 2\pi r_{\phi_{\pi_1}},
 \cos\theta_{\pi_1} = -1+ 2r_{\theta_{\pi_1}}\ .
\enea

After the described changes of variables are performed, the phase
 space reads ($i=I\ {\rm or} \ II$) 

\bgea
\kern-25pt
{\dLips}_4(p_1+p_2;q_1,q_2,\pi_1,\pi_2) =  P_{s,i}
 dr_{k_1^2}dr_{Q^2}dr_{\theta_{k_1}}dr_{\phi_{k_1}}
dr_{\theta_{q_1}}dr_{\phi_{q_1}}dr_{\theta_{\pi_1}}dr_{\phi_{\pi_1}}\ ,
\enea
with 
\bgea
\kern-15pt
 P_{s,i} &=&\frac{1}{4(4\pi)^5s} \sqrt{1-\frac{4m_{\pi}^2}{Q^2}}
\sqrt{1-\frac{4m_e^2}{k_1^2}}\lambda^{1/2}(s,Q^2,k_1^2) 
 \left((\sqrt{s}-2m_e)^2-4m_{\pi}^2\right) \nn \\  && \cdot
\frac{|\vec{k_1}||\vec{p_1}|}{2k_{10}p_{10}-k_1^2}
\left(\frac{-k_1^2+2k_{10}p_{10}}{2|\vec{k_1}||\vec{p_1}|}
 -\cos\theta_{k_1}\right)
\left(\frac{-k_1^2+2k_{10}p_{10}}{2|\vec{k_1}||\vec{p_1}|}
 +\cos\theta_{k_1}\right) \nn
 \\  && \cdot
 P_i  \cdot \ln\left(\frac{-k_1^2+2k_{10}p_{10}+2|\vec{k_1}||\vec{p_1}|}
{-k_1^2+2k_{10}p_{10}-2|\vec{k_1}||\vec{p_1}|}\right)^2 \ ,
\enea
where 
\bgea
 P_I &=& \ln^3\left(\left(1-\sqrt{Q^2/s}\right)^2\right)-\ln^3({4m_e^2/s})
 \ , \\
  P_{II} &=&\ln\left(\frac{(s-4m_e^2)}{\sqrt{Q^2}(2\sqrt{s}-\sqrt{Q^2})}\right)
\enea
for the change of variables from Eq.(\ref{log1}) or Eq.(\ref{log2})
 respectively. Again $0<r_i<1$ for $i=k_1^2,\cdots,{\phi_{\pi_1}}$. 

%%%%%%%%%%%%%%%%%%%%%%%%%%%%%%%%%%%%%%%%%%%%%%%%%%%%%%
\subsection{Merging $s-$ and $t-$channel}

 The function, which approximates the peaking behavior of the
 matrix element reads

 \bgea
 F= \left(\frac{1}{P(q_1,q_2)}+\frac{1}{P(q_2,q_1)}
  +\frac{a}{P_s}\right)^{-1} \ , \ {\rm with} \ 
P_s = \frac{P_{s,I}+\mathrm{b}P_{s,II}}
{\frac{3\ln^2(k_1^2/s)}{k_1^2}+\frac{\mathrm{b}}{s-k_1^2}} \ .
 \enea 

Similarly to the one pion case the cross section 
 Eq.(\ref{eq:master-formula-cs-4}) is  rewritten as
\bgea
\label{eq:master-formula-cs-4a}
\dd\sigma_{avg}(e^+ e^- \to \pi^+ \pi^- e^+ e^-) &=& \frac{1}{4}
\frac{1}{2s}   \; \sum{\left| \mathcal{M}_{\pi^+ \pi^-} \right|^2}(2+a)F
 \times \biggl\{ \nonumber \\
&& \kern -160pt \int_0^A \dd r_9 
dr_{Q^2}dr_{Q'^2}dr_{\theta_{q_2}}dr_{\phi_{q_2}}dr_{\theta_Q}dr_{\phi_Q}
dr_{\theta_{\pi_1}}dr_{\phi_{\pi_1}}
\nonumber \\
 && \kern -180pt  + \int_A^{2A} \dd r_9 
dr_{Q^2}dr_{Q'^2}dr_{\theta_{q_1}}dr_{\phi_{q_1}}dr_{\theta_Q}dr_{\phi_Q}
dr_{\theta_{\pi_1}}dr_{\phi_{\pi_1}}
\nonumber \\
&& \kern -180pt   + \int_{2A}^1 \dd r_9 \times \biggl[
\int_0^B \dd r_{10} +\int_B^1 \dd r_{10}\biggr] dr_{k_1^2}dr_{Q^2}dr_{\theta_{k_1}}dr_{\phi_{k_1}}
dr_{\theta_{q_1}}dr_{\phi_{q_1}}dr_{\theta_{\pi_1}}dr_{\phi_{\pi_1}}\biggr\}\
 .
 \nonumber \\
\enea
with $A=\frac{1}{2+a}$ and $B=\frac{P_I}{P_I+bP_{II}}$. There are
 three channels generated with probabilities $A$,$A$ and $1-2A$, while
 the third channel branches into two channels generated with
 probabilities $B$ and $1-B$. In the first two channels the change of
 variables described in Section \ref{pi2t} is used, while in the third one   
 (branched into two channels) the changes of variables are described
 in Section \ref{pi2s}.
The introduced a priori weights $a$ and $b$, which 
 give the best efficiency of the generation at the $\phi$ meson factory
  were found to be $a=1.1$ and $b=1000$.

\begin{figure} \begin{center}
 \resizebox{0.96\textwidth}{!}{%
      \includegraphics{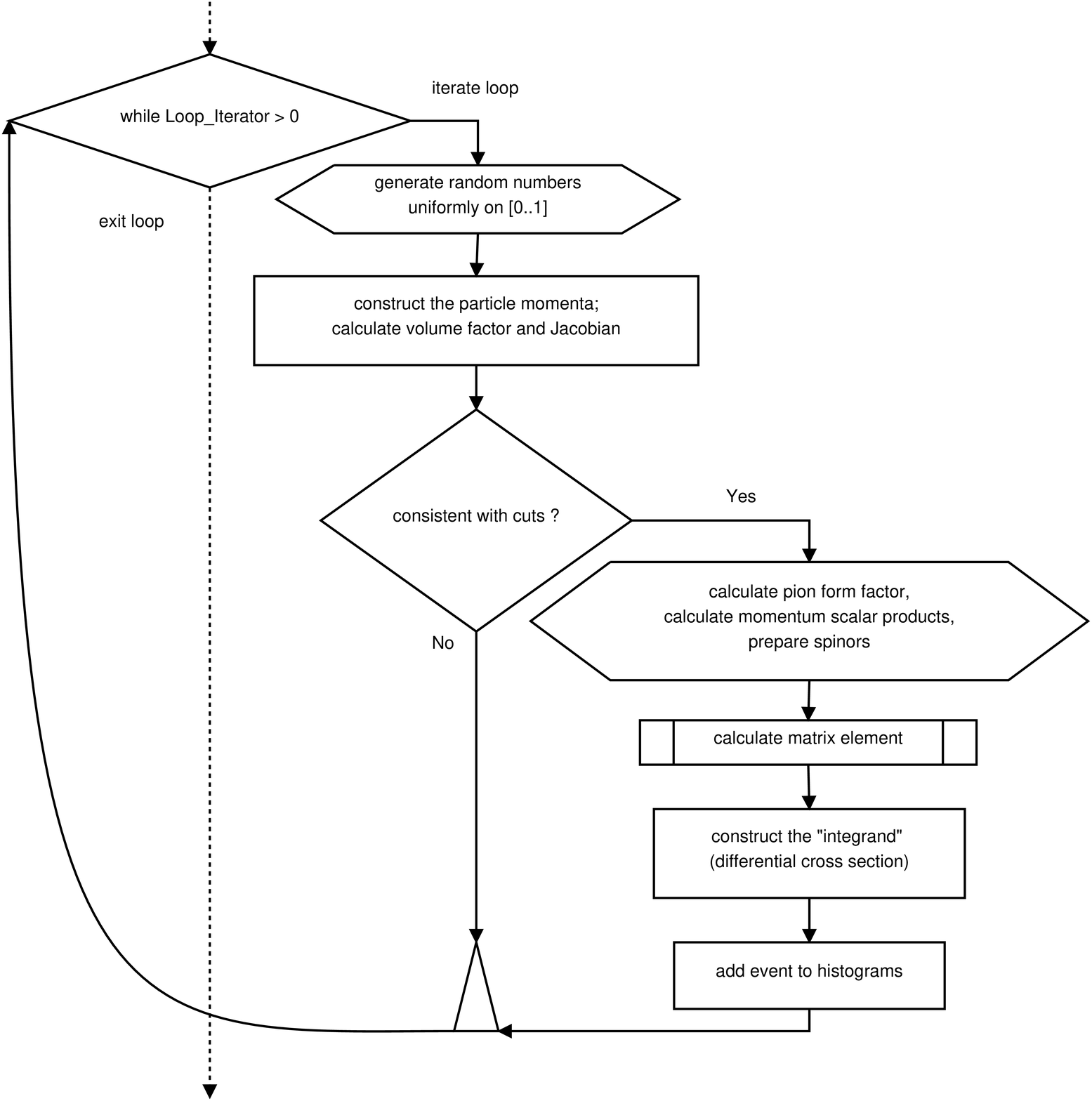} } 
 \end{center}
 \caption{ Flowchart for $e^+e^- \to e^+e^- \pi^+\pi^-$ event generation.
           % {\tt SUBROUTINE mc\_loop\_2pi} (mode $e^+e^- \to e^+e^- \pi^+\pi^-$).
         }
 \label{fig:chart-4}
 \end{figure}

The flowchart for Monte Carlo generator routine is
given in Figure~\ref{fig:chart-4}.
%%%%%%%%%%%%%%%%%%%%%%%%%%%%%%%%%%%%%%%%%%%%%%%%%%%%%%%%%%%%%%%%%%%%%%%%%
%%%%%%%%%%%%%%%%%%%%%%%%%%%%%%%%%%%%%%%%%%%%%%%%%%%%%%%%%%%%%%%%%%%%%%%%%

%%%%%%%%%%%%%%%%%%%%%%%%%%%%%%%%%%%%%%%%%%%%%%%%%%%%%%
\section{Overview of the software structure}
\label{sec:Overview}
 
Let us overview the directory structure of the distribution.
EKHARA is distributed as a source code.
The code of the Monte Carlo generator is located in the
directory {\tt ekhara-routines}. 
The main source file of EKHARA is {\tt ekhara.for}.
There are other source files in the directory {\tt ekhara-routines},
which are automatically included:
\begin{itemize}
 \item%[] 
the $e^+e^- \to e^+e^- \pi^0$ mode is
implemented in
     {\tt routines\_1pi.inc.for}             
and its supplementary histogramming routines are given in
{\tt routines-histograms\_1pi.inc.for};
 \item%[] 
the $e^+e^- \to e^+e^- \pi^+\pi^-$  mode
is coded in 
{\tt routines\_2pi.inc.for},
its supplementary histogramming routines are in
{\tt routines-histograms\_2pi.inc.for}
and helicity-amplitude routines are given in
{\tt routines-helicity-aux.inc.for};
 \item%[] 
 the routines for the matrix and vector manipulations
 are located in {\tt routines-math.inc.for};
 \item%[] 
 in {\tt routines-user.inc.for} 
 several routines are collected, which can be 
 changed by a user in order to customise the operation of EKHARA;
 these include the data-card reading, 
 reporting of events, form factor formulae, filling the histograms,
 additional phase space cuts, etc.; 
 \item%[] 
 all common blocks are  included from the
 file {\tt common.ekhara.inc.for}.
This file contains the detailed comments 
on the explicit purpose 
of the most important common variables.
\end{itemize}

The operation of the EKHARA generator requires the following steps:
\begin{enumerate}
  \item initialization,
  \item event generation,
  \item finalization.
\end{enumerate}

This structure allows us to build a stand-alone generator,
as well as an interface to a separate program (e.g., a detector simulation)
when EKHARA is called on an event-by-event basis.
In the main directory of the distribution there are examples 
of both uses of EKHARA.
A stand-alone example is given in {\tt ekhara-standalone.for}.
An example of an EKHARA interface on an event-by-event basis
is given in {\tt ekhara-call-example.for}.
The main directory of the distributed version contains a {\tt readme.txt} file
with a short description how to compile, run and test the program in the
regimes, described above. 
It is suggested to use the {\tt Makefile}, which is placed
in the main directory.
An example of the full set of input files and the plotting environment is
supplied in the {\tt Env} subdirectory. 
If one uses the distributed {\tt Makefile}, the content of
the {\tt Env} subdirectory will be put into the {\tt EXE} subdirectory
together with an executable {\tt ekhara.exe}
(for details, see Section~\ref{sec:Install} and {\tt readme.txt}).
In the following we assume that {\tt ekhara.exe}
is located  for execution together with the input files in 
the {\tt EXE} subdirectory.

A source code of {\tt RANLUX} random number 
generator\footnote{The unpublished double
precision version of fast RANLUX~\cite{Luscher:1993dy,James:1993vv}
 written by M.~L\"uscher (F.~James, private communication).}
 written in C ({\tt ranlxd.c}), 
together with its C--FORTRAN wrap 
({\tt ranlux\_fort.c} for standard build
 and {\tt ranlux\_fort\_vs.c} for build with {\tt cl}
 in MS~Windows and {\tt xcl} in IBM AIX) 
are supplied in the directory {\tt ranlux-routines}. 

%%%%%%%%%%%%%%%%%%%%%%%%%%%%%%%%%%%%%%%%%%%%%%%%%%%%%%
\subsection{I/O scheme}
\label{subsec:io}

All the input files of EKHARA are supposed to
be located in the same directory as the main executable, {\tt ekhara.exe}.
There are the following types of the input files:
random seeds, parameter input, data-cards and histogram settings.
An example of the full set of input files 
can be found in the {\tt Env} directory.

All the output files of EKHARA 
are written into {\tt ./output} subdirectory.
There are the following types of the output files:
logs of execution, histograms and events.

%%%%%%%%%%%%%%%%%%%%%%%%%%%%%%%%%%%%%%%%%%%%%%%%%%%%%%
\subsubsection*{Input files}

The main input file is called {\tt input.dat}. It contains all global settings:
\begin{itemize}
\item          number of generated events
\item          mode selection (one pion or two pions in the final state)
\item          histogram writing switch
\item          events writing switch
\item          particle masses and constants
\end{itemize}
Another parameter, the random seed mode switch, chooses the way the seeds
for the random number generator are handled.
In the constant seed mode, the file {\tt seed.dat} is used 
for every execution and it is never modified.
In the variable seed mode, the file {\tt seed-v.dat}
is used and on successful completion of each run, a new random
seed is written into {\tt seed-v.dat}. The latter mode
is convenient for a subsequent production of statistically independent samples.

The channel-dependent parameters, which are supposed to be often changed
by a user are collected in ``data-cards''  {\tt card\_1pi.dat} and {\tt card\_2pi.dat}.
These data-cards allow to set the total energy, types of included amplitudes 
and kinematic cuts. A detailed description can be found in comments
within these files.
In {\tt card\_1pi.dat} one can also use the {\tt piggFFsw} switch
in order to select the form of the two photon pion form factor.
\begin{itemize}
   \item[1] (WZWconst) constant form factor; this is not physical
             and should only be used for tests;
   \item[2,3,4] (rho pole, LMD, LMD+V) the form factors given in~\cite{Knecht:2001qf};
   \item[5]  (LMD+V new) the form factor of the lowest meson dominance model with two vector resonances 
              fitted~\cite{Nyffeler:2009uw} to the BABAR data~\cite{:2009mc}.
              This is the recommended form factor;
   \item[6]  (quark) the form factor given in~\cite{Dorokhov:2009jd}.
\end{itemize}
The channel-dependent histogramming settings are given 
in the files   {\tt histo-settings\_1pi.dat} and {\tt histo-settings\_2pi.dat}.

%%%%%%%%%%%%%%%%%%%%%%%%%%%%%%%%%%%%%%%%%%%%%%%%%%%%%%
\subsubsection*{Output: logging}
The main execution log file is the {\tt output/runflow.log}.
It contains the main information about the operation mode and status
of EKHARA, this information is also partly written into the standard
output (i.e., the console).
At the end of a successful execution, the total cross section
is reported to {\tt output/runflow.log} and also to 
the standard output.

A non-standard behaviour of the MC generator
is reported into {\tt output/warnings.log},
while the critical problems in the event generator operation
are reported into the file {\tt output/errors.log}.

In the case of a correct operation, the {\tt output/errors.log} 
and {\tt output/warnings.log} should remain blank.
We strongly recommend to keep track on this issue and report to the
 authors any warnings or errors. 

%%%%%%%%%%%%%%%%%%%%%%%%%%%%%%%%%%%%%%%%%%%%%%%%%%%%%%
\subsubsection*{Output: histograms. Plotting scripts}

When histogramming is allowed through settings in the 
{\tt input.dat}, the plain text files with
the histogram data are saved at the end of the generator
execution.

\begin{itemize}
 \item In $e^+e^- \to e^+e^- \pi^+\pi^-$ mode the file
       {\tt histograms\_2pi.out} contains
       the data for $d \sigma / dQ^2$ histogram.
       One may use the plotting script {\tt doplots.sh}
       from directory {\tt histo-plotting\_2pi}
       in order to plot this histogram 
       (an installed {\tt Gnuplot} is required). 

 \item In $e^+e^- \to e^+e^- \pi^0$ mode there is a wide set
       of histograms stored in the files
       {\tt histo<Number>.<variable>.dat},
       where {\tt<Number>} stands for the histogram number
       and {\tt <variable>} is the histogramming variable acronym.

       One may use the plotting script {\tt do-everything.sh}
       in the directory {\tt histo-plotting\_1pi}
       in order to plot all the histograms and collect them
       into a single postscript file.
       An installed \LaTeX \ system is required for the latter. 
       
       One may use the plotting script {\tt doplots.sh}
       in the directory {\tt t1-t2-bars\_1pi}
       in order to plot the 3D-bar graph, which shows
       the distribution in two variables: $t_1$ and $t_2$.

\end{itemize}
As the histograms are stored as plain text files
the user can use also her/his favourite plotting programs 
to visualize the histograms.

%%%%%%%%%%%%%%%%%%%%%%%%%%%%%%%%%%%%%%%%%%%%%%%%%%%%%%
\subsubsection*{Output: events}

The generated four-momenta of the particles are stored in the following variables,
which can be accessed through common blocks:
\noindent
\begin{longtable}{p{0.3\textwidth} p{0.7\textwidth}}
{\tt p1}&initial positron, 
\\
{\tt p2}&initial electron,
\\
{\tt q1}&final positron,
\\
{\tt q2}&final electron,
\\
{\tt qpion}&final pseudoscalar ($e^+e^- \to e^+e^- \pi^0$ mode),
\\
{\tt pi1, pi2}&final pseudoscalars ($e^+e^- \to e^+e^- \pi^+\pi^-$ mode).
\end{longtable}
In the file {\tt ekhara-call-example.for} we give an example 
how the generated momenta can be used, when EKHARA works
in the event-by-event regime.
In the standalone regime we suggest to 
use the routine {\tt reportevent\_1pi} defined in the file 
{\tt routines-user.inc.for}, which is called
automatically for every accepted unweighted event
($e^+e^- \to e^+e^- \pi^0$ mode only). 
In the distributed version this routine writes the
events to the file {\tt output/events.out}
 when {\tt WriteEvents} flag is on.
One can modify this routine 
in order to accommodate the way how 
the generated events are collected.

%%%%%%%%%%%%%%%%%%%%%%%%%%%%%%%%%%%%%%%%%%%%%%%%%%%%%%
\subsection{Selected procedures}

The top-level interface to the Monte Carlo generator is provided 
by the routine
\noindent
\begin{longtable}{p{0.3\textwidth} p{0.7\textwidth}}
{\tt EKHARA(i) }& \parbox[t]{0.7\textwidth}{
               {\tt i = -1}: initialize, \\
               {\tt i = 0}: generate event(s), \\
               {\tt i = 1}: finalize.
               }
\end{longtable}
\noindent
Only this routine should be called from an external program,
when you use EKHARA in the event-by-event regime,
see example {\tt ekhara-call-example.for}.

In order to  describe briefly the ``internal'' structure
of EKHARA, we list several important routines.

\noindent
\begin{longtable}{p{0.3\textwidth} p{0.7\textwidth}}

{\tt EKHARA\_INIT\_read}  & \parbox[t]{0.7\textwidth}{reading the input files and datacards,}
\\
{\tt EKHARA\_INIT\_set}   & \parbox[t]{0.7\textwidth}{initialization of the MC loop and mappings,}
\\
{\tt EKHARA\_RUN }        & \parbox[t]{0.7\textwidth}{MC loop execution,}
\\
{\tt EKHARA\_FIN}         & \parbox[t]{0.7\textwidth}{MC finalization, saving the results.}
\end{longtable}

%%%%%%%%%%%%%%%%%%%%%%%%%%%%%%%%%%%%%%%%%%%%%%%%%%%%%%
\subsubsection*{ $e^+e^- \to e^+e^- \pi^0$ procedures}

\noindent
\begin{longtable}{p{0.3\textwidth} p{0.7\textwidth}}
{\tt mc\_loop\_1pi}    & \parbox[t]{0.7\textwidth}{
                         Monte Carlo loop
                         (see the flowchart in Figure~\ref{fig:chart-3}),}
\\
{\tt EvalUpperBound}   & \parbox[t]{0.7\textwidth}{evaluation of the upper bound for the Monte Carlo integrand,}
\\
{\tt phasespace\_1pi}  & \parbox[t]{0.7\textwidth}{a wrap for the phase space generation routines,}
\\
{\tt eventselection\_1pi}  & \parbox[t]{0.7\textwidth}{kinematic cuts, }
\\
{\tt m\_el\_1pi}  & \parbox[t]{0.7\textwidth}{calculation of the matrix element for $e^+e^- \to e^+e^- \pi^0$,}
\\
{\tt  tellSIGMA\_1pi}  & \parbox[t]{0.7\textwidth}{reports the total cross section.}
\end{longtable}

%%%%%%%%%%%%%%%%%%%%%%%%%%%%%%%%%%%%%%%%%%%%%%%%%%%%%%
\subsubsection*{ $e^+e^- \to e^+e^- \pi^+\pi^-$ procedures}

\noindent
\begin{longtable}{p{0.3\textwidth} p{0.7\textwidth}}
{\tt mc\_loop\_2pi}    & \parbox[t]{0.7\textwidth}{
                         Monte Carlo loop
                         (see the flowchart in Figure~\ref{fig:chart-4}),}
\\
{\tt eventselection\_2pi}  & \parbox[t]{0.7\textwidth}{kinematic cuts,}
\\
{\tt  phsp1}  &\parbox[t]{0.7\textwidth}{phase space generation routine (branch 1),}
\\
{\tt  phsp2}  &\parbox[t]{0.7\textwidth}{phase space generation routine (branch 2),}
\\
{\tt  phsp3}  &\parbox[t]{0.7\textwidth}{phase space generation routine (branch 3),}
\\
{\tt  matrixelmt}  &\parbox[t]{0.7\textwidth}{ calculation of the matrix element for $e^+e^- \to e^+e^- \pi^+\pi^-$,}
\\
{\tt  tellSIGMA\_2pi}  &\parbox[t]{0.7\textwidth}{ reports the total cross section.}
\end{longtable}
For details see the comments in the source code.

%%%%%%%%%%%%%%%%%%%%%%%%%%%%%%%%%%%%%%%%%%%%%%%%%%%%%%
\section{Compilation instructions}
%\section{Installation instructions}
\label{sec:Install}

Being distributed as a source code the program
does not require installation, but a compilation and linking is needed.
EKHARA does not need any specific external libraries.
In order to compile the program, a user 
may run any OS (UNIX, Linux, MS Windows, etc)
with correctly installed 
\begin{itemize}
 \item FORTRAN 77 compiler with support of quadruple precision,
 \item C compiler.
\end{itemize}
The program was tested on the following platforms:
\begin{itemize}
 \item Linux (Ubuntu 8.04.3)\\
       GNU C Compiler (gcc) 4.2.4 \\
       Intel(R) FORTRAN Compiler (ifort) 11.0.20080930
 \item MS Windows (XP SP3) \\
       MS Visual Studio 2008 (nmake, cl) \\
       Intel(R) FORTRAN Compiler (ifort) 11.
\end{itemize}
The program distribution contains the {\tt Makefile},
with targets for Linux, IBM AIX and Windows 
environments.
The {\tt Makefile} is annotated,
in order to help a user to tune it up for the
 own requirements.
The main {\tt Makefile} targets are listed in  
Table~\ref{tabmakefile}.

\begin{table}
\begin{longtable}{p{0.4\textwidth}  p{0.16\textwidth} p{0.17\textwidth} p{0.17\textwidth}}
\hline
Description & Linux & Windows & IBM AIX
\\
\hline
\vspace{0.008\baselineskip}
\\
\parbox[t]{0.4\textwidth}{ Default: Standalone MC generator, sample input files and
                           histogramming routines. Everything put into 
                           {\tt EXE} directory
                         }
                         & {\tt default} &  {\tt default-vs} & {\tt default-ibm}
\\
\vspace{0.008\baselineskip}
\\
\parbox[t]{0.4\textwidth}{ Compile everything: default, ranlux-testing program and 
                           seed-production programs
                         }
                         & {\tt all} &  {\tt all-vs} & {\tt all-ibm}
\\
\vspace{0.008\baselineskip}
\\
\parbox[t]{0.4\textwidth}{ Testrun: Compile everything and execute the testrun scripts
                         }
                         & {\tt test} &  {\tt test-vs} & {\tt test-ibm}
\\
\vspace{0.008\baselineskip}
\\
\parbox[t]{0.4\textwidth}{ Remove the redundant and temporary files
                         }
                         & {\tt clean} &  {\tt clean-vs} & {\tt clean-ibm}
\\
\vspace{0.008\baselineskip}
\\
\hline
\end{longtable}
\caption{The main {\tt Makefile} targets\label{tabmakefile}
}
\end{table}

For example, in Linux, a simple way to compile a program is to issue 
{\tt make default}, being in the directory where the {\tt Makefile} is located.
This will produce {\tt ekhara.exe} (main program executable) and copy it
into the sub-directory {\tt EXE}, together with the contents of 
{\tt Env} sub-directory.
The latter contains the set of sample input files and 
histogram plotting scripts.
We provide a full set of necessary 
input files in the distribution package.
It is advised to execute {\tt ekhara.exe}
in the directory {\tt EXE},
where it is placed by default.
Every time one does {\tt make default}, the input files 
in  {\tt EXE} are replaced with the sample ones from {\tt Env}.

In order to produce only an object file with the
EKHARA MC generator, one can use, for example
\begin{verbatim}
	ifort -c ekhara-routines/ekhara.for -o ./ekhara.o
\end{verbatim}

EKHARA needs a random seed for operation.
Different random seeds can be obtained
by using a {\tt Makefile} target {\tt seed\_prod-ifort}.
It produces an exacutable program {\tt seed\_prod.exe},
which generates a set of random seeds.

%%%%%%%%%%%%%%%%%%%%%%%%%%%%%%%%%%%%%%%%%%%%%%%%%%%%%%
\section{Test run description}
\label{sec:Testrun}

It is recommended to test the random 
number generator on a given machine,
before using EKHARA.
It is also important to check whether 
EKHARA can function properly on a given operational system
and that there are no critical bugs due to 
the compiler.
We provide a testrun package for these purposes.

It is suggested to use the {\tt Makefile} targets
{\tt test}, {\tt test-vs} or {\tt test-ibm} 
depending on your environment (Linux, Windows and IBM AIX,
correspondingly).
This will automatically prepare and execute
the following two test steps.

The first step of the testrun is 
the random number generator control.
The source file {\tt testlxf.for} contains
the {\tt ranlux} test routines.

The second step is the verification
if the user-compiled EKHARA can reproduce
the set of results, created by a well-tested copy 
of EKHARA in various modes.
The testrun environment contains directories 
{\tt test} and {\tt test-vs} with precalculated data for a comparison,
along with the random seed and input files for each mode. 
The script {\tt test.sh} is responsible for the 
execution of a user-compiled {\tt ekhara.exe} in all 
the control modes and for the comparison of the output.

Please read carefully the output of the testrun execution
in your console and be sure there are no warnings and/or error
messages.

%%%%%%%%%%%%%%%%%%%%%%%%%%%%%%%%%%%%%%%%%%%%%%%%%%%%%%
\section{Customization of the source code by a user}
\label{sec:User}

We leave for a user an option to customise the generator to
her/his needs by editing the source code file 
{\tt ekhara-routines/routines-user.inc.for}.
Notice, we always use explicit declaration of identifiers and
the {\tt implicit none} statement is written down in each routine.

In the file {\tt ekhara-routines/routines-user.inc.for}
one can change 
\begin{itemize}
\item the data-card reading
(routines {\tt read\_card\_1pi} and {\tt read\_card\_2pi}),
\item the form-factor formula
(routine {\tt piggFF}),
\item the events reporting (routine {\tt reportevent\_1pi}),
\item histogramming (routines {\tt histo\_event\_1pi} and {\tt histo\_event\_2pi}),
\item additional kinematic cuts
(routines {\tt ExtraCuts\_1pi} and {\tt ExtraCuts\_2pi}).
\end{itemize}

%%%%%%%%%%%%%%%%%%%%%%%%%%%%%%%%%%%%%%%%%%%%%%%%%%%%%%
\section{Validation of the generator}
\label{sec:Validation}
In Fig.~\ref{fig:MC-experiment-compar}
we demonstrated an agreement of the Monte Carlo simulation 
of $e^+e^- \to e^+e^- \pi^0$ in the 
``single-tag'' mode with the experimental data from
CLEO~\cite{Gronberg:1997fj} and BaBar~\cite{:2009mc}.
We conclude that the matrix element is well under control
and the applied pion transition form factor 
(LMD+V)~\cite{Nyffeler:2009uw} is in agreement with data.
The procedure of the phase space generation was validated
by means of the high statistics phase space volume calculation
in EKHARA and comparison of the result with that from 
 the independent dedicated numerical calculation. 
The volume was also compared to that obtained by 
GALUGA generator~\cite{Schuler:1997ex}.
We have also verified that our numerical results 
for the three-body phase space volume 
in the limit of massless $\pi^0$ reproduce
well those of the analytic expression.
 
The $e^+e^- \to e^+e^- \pi^+\pi^-$ mode is validated
by means of the reproduction of the results from 
the previous version of EKHARA.
In~\cite{Czyz:2005ab,Czyz:2006dm,Czyz:2003gb} the tests of this part
are presented in detail.

%%%%%%%%%%%%%%%%%%%%%%%%%%%%%%%%%%%%%%%%%%%%%%%%%%%%%%
\section{Summary}
\label{sec:Conclusion}
 An update (version 2.0) of the  Monte Carlo event generator
 EKHARA is presented.
  It generates  processes
  $e^+e^- \to e^+e^- \pi^0$ and $e^+e^- \to e^+e^- \pi^+\pi^-$.
 The newly added channel ($e^+e^- \to e^+e^- \pi^0$)  is
 important for $\gamma^{*}\gamma^{*}$ physics and
 can be used for the pion transition form factor 
 studies at meson factories.

%%%%%%%%%%%%%%%%%%%%%%%%%%%%%%%%%%%%%%%%%%%%%%%%%%%%%%
\subsection*{Acknowledgements}
\label{sec:Acknowl}

We would like to thank 
Fred Jegerlehner and Andreas Nyffeler for discussion of the physics case
of $\gamma^*\gamma^*\to\pi^0$,
Vladimir Druzhinin for drawing our attention to Ref.~\cite{Schuler:1997ex},
Danilo Babusci, Dario Moricciani and Graziano Venanzoni 
for discussion of the experimental project KLOE-2 
and simulation issues.
We are grateful to Germ\'an Rodrigo for his kind hospitality
at Instituto de F\'isica Corpuscular CSIC (Valencia)
and to Achim Denig at Institut f\"ur Kernphysik  J.Gutenberg-Universit\"at (Mainz), 
where a part of this work has been done.
This work was partially
 supported by MRTN-CT-2006-035482 ``FLAVIAnet''
 under the Sixth Framework Program of EU,
Polish Ministry of Science and High Education
   from budget for science for years 2010-2013: grant number N N202 102638 and 
the European Community-Research 
Infrastructure Integrating Activity 
``Study of Strongly Interacting Matter'' 
(acronym HadronPhysics2, Grant Agreement n. 227431)
 under the Seventh Framework Program of EU.

%%%%%%%%%%%%%%%%%%%%%%%%%%%%%%%%%%%%%%%%%%%%%%%%%%%%%%%%%%%


%\appendix
%\section{Appendices}
%\label{sec:App}
%% 
%%%%%%%%%%%%%%%%%%%%%%%%%%%%%%%%%%%%%%%%%%%%%%%%%%%%%%%%%%%
%\subsubsection*{}

%%%%%%%%%%%%%%%%%%%%%%%%%%%%%%%%%%%%%%%%%%%%%%%%%%%%%%%%%%%
%%%%%%%%%%%%%%%%%%%%%%%%%%%%%%%%%%%%%%%%%%%%%%%%%%%%%%%%%%%
%% References


\bibliographystyle{model1-num-names}

\bibliography{ekhara-preprint}      %%% S.I. > for arXiv

%\bibliography{ivashyn}      %%% S.I. > for my latex on Ubuntu
%\bibliography{ivashyn.bib} %%% S.I. > for my MIKTEX 2.8


\end{document}

%% file: preamble.tex
\usepackage{graphicx}
\usepackage{amssymb}
\usepackage{amsmath,latexsym} %%extra included by S.I.

\usepackage{longtable}

\journal{Computer Physics Communications}

\newcounter{bla}
\newenvironment{refnummer}{%
\list{[\arabic{bla}]}%
{\usecounter{bla}%
 \setlength{\itemindent}{0pt}%
 \setlength{\topsep}{0pt}%
 \setlength{\itemsep}{0pt}%
 \setlength{\labelsep}{2pt}%
 \setlength{\listparindent}{0pt}%
 \settowidth{\labelwidth}{[9]}%
 \setlength{\leftmargin}{\labelwidth}%
 \addtolength{\leftmargin}{\labelsep}%
 \setlength{\rightmargin}{0pt}}}
 {\endlist}

\def\bge{\begin{equation}}
\def\ene{\end{equation}}
\def\bgea{\begin{eqnarray}}
\def\enea{\end{eqnarray}}
\def\nn{\nonumber}

\def\dd{\mathrm{\,d\,}}
\def\dLips{\mathrm{\,d\,Lips}}

\def\VV{V}

%% file: frontmatter.tex
\begin{frontmatter}



\title{\ourtitle}


\author[label1]{Henryk Czy\.z}

\author[label1,label2]{Sergiy Ivashyn\corref{cor1}}

\address[label1]{Institute of Physics, University of Silesia, Uniwersytecka 4, Katowice PL-40007, Poland}
\address[label2]{NSC ``KIPT'', Akademicheskaya 1, Kharkov UA-61108, Ukraine}
\cortext[cor1]{Corresponding author
              }
\ead{ivashyn@kipt.kharkov.ua}
%\address[label1]{Institute of Physics, University of Silesia, Katowice PL-40007, Poland}
%\address[label2]{NSC ``KIPT'', Kharkov UA-61108, Ukraine}

\begin{abstract}
  We present EKHARA Monte Carlo event generator of reactions
  $e^+e^- \to e^+e^- \pi^0$ and $e^+e^- \to e^+e^- \pi^+\pi^-$.
 The newly added channel ($e^+e^- \to e^+e^- \pi^0$)  is
 important for $\gamma^{*}\gamma^{*}$ physics and
 can be used for the pion transition form factor 
 studies at meson factories.

% Moreover its better
% experimental knowledge will contribute to the error reduction
% in calculations of the light--by--light
% contributions to the muon anomalous magnetic moment.

\end{abstract}

\begin{keyword}
%% (up to six Keywords!!)
{Monte Carlo generator} \sep 
{Pion transition form factor}  \sep 
{Pion pair production}  \sep 
{Two-photon processes}  \sep 
{$e^+ e^-$ annihilation}
\end{keyword}

\end{frontmatter}

%% file: prog-summary.tex
\begin{small}
\noindent
{\em Manuscript Title:} \ourtitle
                                      \\
{\em Authors:} H.~Czy\.z, S.~Ivashyn
                                               \\
{\em Program Title:} EKHARA
                                         \\
{\em Journal Reference:}                                      \\
  %Leave blank, supplied by Elsevier.
{\em Catalogue identifier:}                                   \\
  %Leave blank, supplied by Elsevier.
{\em Licensing provisions:} none
                                   \\
  %enter "none" if CPC non-profit use license is sufficient.
{\em Programming language:} FORTRAN 77
                            with quadruple precision
                                  \\
{\em Computer:} PC, main frame
                                              \\
  %Computer(s) for which program has been designed.
{\em Operating system:} Linux, Unix, MS Windows
                                      \\
  %Operating system(s) for which program has been designed.
{\em RAM:} up to 10 Megabytes for operation of the compiled program
                                             \\
  %RAM in bytes required to execute program with typical data.
{\em Number of processors used:} one
                             \\
  %If more than one processor.
% {\em Supplementary material:}                                 \\
%   % Fill in if necessary, otherwise leave out.
{\em Keywords:} 
  % Please give some freely chosen keywords that we can use in a
  % cumulative keyword index. (up to six Keywords!!)
{Monte Carlo generator}, \sep 
{Event simulation}, \sep 
{Pion transition form factor},  \sep 
{Pair production},  \sep 
{Two-photon process},  \sep
{e+ e- collision}
 \\
{\em Classification:} 11.2 Phase Space and Event Simulation,
11.6 	Phenomenological and Empirical Models and Theories
                                        \\
  %Classify using CPC Program Library Subject Index, see (
  % http://cpc.cs.qub.ac.uk/subjectIndex/SUBJECT_index.html)
  %e.g. 4.4 Feynman diagrams, 5 Computer Algebra.
% {\em External routines/libraries:}                                      \\
%   % Fill in if necessary, otherwise leave out.
% {\em Subprograms used:}                                       \\
%   %Fill in if necessary, otherwise leave out.
%
{\em Nature of problem:}\\
  %Describe the nature of the problem here.
   The first version of EKHARA [1,2] 
  was developed to simulate background for the  
  pion form factor measurement at meson factories coming from
  the process  $e^+ e^- \to e^+ e^- \pi^+ \pi^-$. The newly added
 channel $e^+ e^- \to e^+ e^- \pi^0$  will help in the pion transition
 from factor studies at meson factories [3].
   \\
{\em Solution method:}\\
Events consisting of the momenta of the 
outgoing particles are generated by Monte Carlo methods.
The generated events are distributed accordingly 
to the theoretical cross section. For the $e^+ e^- \to e^+ e^- \pi^0$
mode the Monte Carlo sampling developed in [4] was adopted. 
  %Describe the method solution here.
   \\
{\em Restrictions:}\\
  %Describe any restrictions on the complexity of the problem here.
  In order to compile the code, 
  the FORTRAN 77 compiler should support quadruple precision numbers.
   \\
{\em Unusual features:}\\
 Calculations are carried in quadruple precision,
 in order to avoid numerical cancellations
 in $e^+ e^- \to e^+ e^- \pi^+ \pi^-$ mode.
  %Describe any unusual features of the program/problem here.
   \\
%{\em Additional comments:}\\
  %Provide any additional comments here.
   \\
{\em Running time:}\\
  Depends on the requested mode and applied kinematic cuts. 
  Example: on Intel Core2 Quad CPU Q6600 @ 2.40GHz,
  using only one thread, 
  \begin{itemize}
    \item 
  $10^5$ unweighted $e^+e^-\pi^0$ events are generated 
  in $78$~seconds (no cuts),
    \item 
  $10^5$ weighted $e^+e^-\pi^+\pi^-$ events are generated 
  in $38$~seconds (with cuts~[2]).
  \end{itemize}
   %\\
{\em References:}
\begin{refnummer}
\item  H.~Czyz, E.~Nowak-Kubat, 
       Radiative return via electron pair production: 
        Monte Carlo simulation of the process 
        $e^+ e^- \to \pi^+ \pi^- e^+ e^-$,
        Acta Phys. Polon., 2005, B36, 3425--3434
\item   H.~Czyz, E.~Nowak-Kubat, 
      The reaction $e^+ e^- \to e^+ e^- \pi^+ \pi^-$ and
       the pion form factor measurements via the radiative return method,
       Phys. Lett., 2006, B634, 493--497
\item G.~Amelino-Camelia and others,
          Physics with the KLOE-2 experiment at the upgraded DA$\Phi$NE,
          Eur.Phys. J., 2010, C68, 619--681
\item G.A.~Schuler, 
      Two-photon physics with GALUGA 2.0,
      Comput. Phys. Commun., 1998, 108, 279--303

\end{refnummer}
\end{small}